# Observation of topological transition in high-$T_c$ superconductor FeTe$_{1-x}$Se$_x$/SrTiO$_3$(001) monolayers

X.-L. Peng[1,2], Y. Li[1,2], X.-X. Wu[3], H.-B. Deng[1,2], X. Shi[1,2], W.-H. Fan[1,2], M. Li[4], Y.-B. Huang[4], T. Qian[1,5,6], P. Richard[7], J.-P. Hu[1,2,5,6], S.-H. Pan[1,2,5,6], H.-Q. Mao[1†], Y.-J. Sun[1,5,6†] & H. Ding[1,2,5,6†]

[1]*Beijing National Laboratory for Condensed Matter Physics, and Institute of Physics, Chinese Academy of Sciences, Beijing 100190, China*

[2]*School of Physics, University of Chinese Academy of Sciences, Beijing 100190, China*

[3]*Institut für Theoretische Physik und Astrophysik, Julius-Maximilians-Universität Würzburg, Würzburg, Germany.*

[4]*Shanghai Synchrotron Radiation Facility, Shanghai Institute of Applied Physics, Chinese Academy of Sciences, Shanghai, China.*

[5]*Songshan Lake Materials Laboratory, Dongguan, Guangdong 523808, China*

[6]*CAS Center for Excellence in Topological Quantum Computation, University of Chinese Academy of Sciences, Beijing 100190, China*

[7]*Institut quantique, Université de Sherbrooke, 2500 boulevard de l'Université, Sherbrooke, Québec J1K 2R1, Canada*

**ABSTRACT**

Superconductors with topological surface or edge states have been intensively explored for the prospect of realizing Majorana bound states, which obey non-Abelian statistics and are crucial for topological quantum computation. The traditional routes for making topological insulator/superconductor and semiconductor/superconductor heterostructures suffer fabrication difficulties and can only work at low temperature. Here, we use angle-resolved photoemission spectroscopy to directly observe the evolution of a topological transition of band structure nearby the Fermi level in two-dimensional high-$T_c$ superconductor FeTe$_{1-x}$Se$_x$/SrTiO$_3$(001) monolayers, fully consistent with our theoretical calculations. Furthermore, evidence of edge states is revealed by scanning tunneling spectroscopy with assistance of theoretical calculations. Our study provides a simple and tunable platform for realizing and manipulating Majorana states at high temperature.

## I. INTRODUCTION

Topological superconductors, which can host Majorana bound states on their surface or edge, have drawn enormous attention in condensed matter physics [1, 2]. This is because the non-Abelian braiding properties of Majorana states are predicted to play an essential role in topological quantum computation. Intrinsic $p$-wave superconductors [3, 4] and heterostructures of $s$-wave superconductors and topological insulators/Rashba-type semiconductors [5, 6] are two typical routes for realizing topological superconductors (or more precisely, superconducting topological states). However, the former suffers from the scarcity of candidate materials and their sensitivity to disorder, whereas the latter suffers from structural complexities and low working temperatures, hindering manipulation of Majorana states in applications. Recently, bulk superconductors with topological surface states [7-16] provide a new route, where ordinary superconductivity and nontrivial topology can be integrated in a single material, such as FeTe$_{0.55}$Se$_{0.45}$ [9-12], Bi$_2$Pd [13] and other iron-based superconductors [14-16]. Among them, the most interesting systems are the high-$T_c$ superconductors, which can significantly raise the working temperature.

Monolayer FeSe grown on SrTiO$_3$(001) (STO) holds the highest $T_c$ (above 65 K [17-24]) among all iron-based superconductors and can potentially combines superconductivity and topological properties into a single material [25-27]. One way of integration is to adjust lattice parameters, for example by substituting selenium by tellurium, which maintains high $T_c$ for monolayer FeTe$_{1-x}$Se$_x$ films within a wide range of substitution, as revealed by previous scanning tunneling microscopy/ spectroscopy (STM/STS) [28] and angle-resolved photoemission spectroscopy (ARPES) results [8]. Moreover, a theoretical work predicted that the monolayer FeTe$_{1-x}$Se$_x$ system can realize topological nontrivial states through band inversion at the Γ point [29] and some supporting evidences were found later in ARPES measurements [8]. However, it is far from settled that this monolayer system can serve as a simple and tunable platform for realizing topological superconductivity at high temperature.

In this paper, we perform systematic ARPES measurements with photon energy and polarization dependence on FeTe$_{1-x}$Se$_x$/STO monolayers with different Se content ($x$) which are grown by *in situ* molecular beam epitaxy (MBE). In addition to the observation of two bands intersecting, the spectral weight and effective mass of bands change dramatically around $x = 0.21$, further supporting a transition of topological band inversion nearby the Fermi level ($E_F$). Furthermore, evidence of edge states of monolayer FeTe/STO around 50 meV below $E_F$ is observed by STM/STS measurements. Their topological origin can be interpreted by our LDA calculations and band structure results.

## II. EXPERIMENTAL AND COMPUTATIONAL METHODS

The crystal structure of monolayer FeTe$_{1-x}$Se$_x$/STO is presented in Fig. 1(a). Monolayer FeTe$_{1-x}$Se$_x$ for *in situ* ARPES measurements are grown on 0.7wt% Nb-doped STO substrates after degassing for 10 hours at 600 °C and then annealing for 1.5 hours at 950 °C in an ultrahigh vacuum MBE chamber. Substrates are kept at 310 °C for monolayer FeSe and 275 °C for monolayer FeTe$_{1-x}$Se$_x$ during the film growth. Fe (99.98%), Se (99.999%) and Te (99.99%) are co-evaporated from Knudsen cells. The flux ratio of Fe to Se/Te is 1:10 which are measured by a quart crystal balance. The growth rate is determined by Fe flux and equaled to 0.7 UC/min. The Se content ($x$) is controlled by the ratio of Te to Se flux speed during growth. After growth, the monolayer FeSe films are annealed at 370 °C and monolayer FeTe$_{1-x}$Se$_x$ at 260 °C for 20 hours. During the growth process, the sample quality is monitored using reflection high-energy electron diffraction (RHEED), as shown in Fig. 1(b). Then the samples are transferred *in situ* to the ARPES chamber for measurements. The exact Se content $x$ of the samples is determined by the *in situ* X-ray photoelectron spectroscopy (XPS) method. The results are shown in Fig. 1(c), where the value of [$x/(1-x)$] of six different samples is proportional to the ratio of the characteristic peak area associated to Se and Te. ARPES measurements are recorded at the BL-09U "Dreamline" beamline of Shanghai Synchrotron Radiation Facility (SSRF), using a VG DA30 electron analyzer under ultrahigh vacuum better than $5 \times 10^{-11}$ torr. The energy resolution is set to ~12 meV for the band structure and ~16 meV for Fermi surface mapping, while the angular resolution is set to 0.2°. Spectra are recorded at 30 K except for special clarification. Growth procedures of samples for *in situ* STM/STS measurements are similar to that introduced above. STM measurements are carried out in ultra-high-vacuum condition at 4.3 K with a home-made STM-MBE combined system. W Tips are carefully calibrated on a Ag island before measurements. The tunneling conductance spectra are obtained using standard lock-in amplifier techniques with a root-mean-square oscillation voltage of 3 meV and a lock-in frequency of 791.1 Hz.

Density functional theory (DFT) calculations employ the projector augmented wave (PAW) method encoded in the Vienna *ab initio* simulation package (VASP) [30-32]. The projector augmented-wave method is used to describe the wavefunctions near the core and the generalized gradient approximation within the Perdew-Burke-Ernzerhof (PBE) parameterization is employed as the electron exchange-correlation functional [33]. For band structure calculations, the cutoff energy of 500 eV is taken for expanding the wave functions into plane-wave basis. In the calculations, the Brillouin zone is sampled in the $k$ space within Monkhorst-Pack scheme [34]. The spin-orbit coupling is taken into account by the second variation method. For 2D free standing FeSe monolayer, the numbers of these $k$ points are $11\times 11$. In the edge states calculations, a 24 Å vacuum layer is adopted and there are 59 Fe atoms and 59 Se atoms for [100] edge and 60 Fe atoms and 62 Se atoms for [110] edge.

### III. RESULTS

The unit cell of monolayer FeTe$_{1-x}$Se$_x$/STO is tetragonal and contains two iron atoms. The in-plane lattice constant $a$ is fixed at 3.905 Å by the STO substrate. The main effect of the isovalent substitution of Te by Se is increasing the height of Se/Te surface $H$, which is labelled in Fig. 1(a). In addition, this substitution can enhance the spin-orbit coupling (SOC) splitting for $d$-orbital bands due to the intrinsic $p$-$d$ hybridization. The Fermi surface of monolayer FeTe$_{1-x}$Se$_x$/STO ($x = 0.19$) is shown in Fig. 1(d). There are electron pockets at the M point, which have been observed in all superconducting monolayer FeTe$_{1-x}$Se$_x$/STO samples. They are believed to play an important role in promoting high-$T_c$ superconductivity in this system [8]. Our local-density approximation (LDA) calculations reveal that a band inversion can be induced around the Γ point of freestanding monolayer FeSe with varying anion height. Note that freestanding monolayer films lack electron doping from substrates compared with samples grown on STO. Figure 1(e) depicts the band structure with a low anion height before band inversion. Around the Γ point, there are three hole-like bands (α, β, γ) and one electron-like band (η). With increasing $H$, the hybridization between Fe $d_{xy}$ and Se $p_z$ weakens and the η band starts to sink. As the η band and the α, β, γ bands have opposite inversion-symmetry eigenvalues, a band inversion between them will drive the system into a topologically nontrivial phase. As shown in Fig. 1(f) for $H = 1.54$ Å, the orbital character change of the η band bottom from $p_z$/$d_{xy}$ to $d_{yz}$/$d_{xz}$ clearly demonstrates this band inversion. Simultaneously, the α band top changes from $d_{xz}$ to $p_z$/$d_{xy}$ and this induces a parity change from even to odd with respect to the xz plane, which is a direct evidence for the band inversion that can be verified by ARPES experiments.

In ARPES measurements, orbital and parity characters of bands can be determined by changing the photon energy and the polarization according to the matrix element effect [35, 36]. We demonstrate the observation of two hole-like bands (α, β) and one electron-like band (η) around the Γ point of monolayer FeTe$_{1-x}$Se$_x$ ($x = 0.21$) in Fig. 2 using photons with different energy and polarizations. Using our experimental facility, as shown in Fig. 2(a), odd (even) orbital with respect to the M$_x$ mirror plane can be observed with $p$- ($s$-) polarized photons. The $d_{yz}$ orbital is odd while $d_{xz}$ is even. Moreover, $p_z$ orbital can be observed by $p$-polarized photons, which also contain the z component of photon polarization (vector potential $A_z$) in our facility geometry. The results of the matrix element effect analysis are summarized in Table 1. Furthermore, to distinguish $p$ and $d$ orbitals, we can take advantage of the photoemission cross section difference of Fe $3d$ and Te $5p$ (Se $4p$) in the energy range from 15 eV to 40 eV [37], as shown in Fig. 2(b).

Figures 2(c)-2(d) display ARPES results and the corresponding second derivative spectrum measured with $p$-polarized, 22 eV photons. The α and η bands can be observed clearly. These two bands form a Dirac-cone-like band structure, indicating that the system is close to the topological transition critical point. The energy of the Dirac point (DP) can be estimated in the momentum distribution curves (MDCs) plot, as shown in Fig. 2(i). When the photon energy increases to 32 eV, as shown in Figs. 2(e)-2(f), the spectral weight of the α and η bands is obviously suppressed, confirming the $p_z$ orbital character in these bands. The residual weak spectral weight of the η band is attributed to its $d_{xy}$ orbital character according to our LDA analysis [Fig. S4 of Supplemental Material [38] ]. On the other hand, the β band becomes more apparent, suggesting its $d_{yz}$ orbital character with odd parity. In Figs. 2(g)-2(h), when we change the polarization geometry from $p$ to $s$, only the α band can be detected, confirming its $d_{xz}$ orbital character with even parity. In the measurements, the third γ band with dominant $d_{xy}$ orbital component can hardly be observed because the spectral weight of $d_{xy}$ is the smallest due to the strongest correlation among all the orbitals in iron-based superconductors [39]. By fitting the experimental bands to $E = C_0 + C_1|k| + C_2 k^2$, we determine that the band top of the α and β bands are located at $21.7 \pm 0.5$ meV and $25.7 \pm 1.2$ meV below $E_F$, respectively, and that the band bottom of η band is at $20.3 \pm 1.2$ meV below $E_F$. We summarize the results of the band structure and orbital analysis of the $x = 0.21$ sample in Fig. 2(j).

If there is a topological transition in monolayer FeTe$_{1-x}$Se$_x$, as predicted in the aforementioned LDA calculations, the orbital and parity characters of the α, β and η bands should exhibit a dramatic change through it. Moreover, as shown above, the sample with Se content $x = 0.21$ is already close to the critical point. To visualize the band inversion process, we systematically study the band structure evolution of monolayer FeTe$_{1-x}$Se$_x$ as a function of the Se content $x$.

Figures 3(a)-3(c) show the band structures of six different Se content samples detected by 22 eV $p$-, 32 eV $p$- and 32 eV $s$-polarized photons. With one particular photon energy, we can observe dramatic changes of bands in samples with different content. In Fig. 3(a), the spectra weight of the α band increases dramatically with decreasing Se content indicating the increases of $p_z$ orbital character in the α band. Moreover, as indicated by a red arrow in Fig. 3(a), the α band top cannot be observed before the band inversion ($x > 0.21$). As the orbital character of the α band top changes to $p_z/d_{xy}$ during the band inversion process, the α band top becomes significantly enhanced after band inversion ($x < 0.21$). The enhancement of α band can also be found in Fig. 3(b), which shows the evolution of $d$-orbital bands with odd parities. The β band is expected to be clear and the α band to be completely suppressed before the band inversion due to its even parity nature. However, the missing spectral weight indicated by a red arrow becomes apparent after the band inversion. Compared with Fig. 3(a), the re-emerging band is attributed to the α band, which clearly suggests that the parity of the α band top changes from even to odd. In contrast to results of $p$-polarized photons, the evolution of the α band exhibits the opposite behavior when measurements are performed with 32 eV $s$-polarized photons and only orbitals with even parities can be detected, as shown in Fig. 3(c). The spectral weight of the α band top is clear indicated by a green arrow before the band inversion and is comparative suppressed after the band inversion, which is also consistent with the above analysis.

We can also trace the evolution of the β band. In contrast to the α band, no obvious change occurs for the β band, except for a band shift to a lower binding energy. To see it more clearly, we show the MDCs plot of the α and β bands in Figs. 3(d)-3(e), respectively. The peak position of the α band continuously evolves toward the Γ point, while that of the β band is almost unchanged with the Se content decreasing. Then, we calculate the effective mass of the α, β and η bands. The results are shown in Fig. 3(f). The effective mass of the three bands exhibits rather different trends with the variation of the Se content. When the Se content decreases, the value of the α band decreases dramatically while that of the η band increases rapidly. However, the effective mass of the β band remains almost unchanged. These changes of effective masses can be interpreted in terms of the orbital character change due to band inversion. Because the *d* orbitals exhibit stronger correlation effect than the *p* orbitals, the η band becomes more correlated due to the increasing *d* orbital components, while the α band behaves oppositely due to the increasing *p* orbital component, and the β band shows nearly no change, in agreement with its relative *d* and *p* components remaining the same in the process of band inversion.

According to the bulk-boundary correspondence, topological edge states (TESs) will appear at the edge of nontrivial monolayer FeTe$_{1-x}$Se$_x$/STO films [1]. We carry out LDA calculations for Fe-Te [100] and Fe-Fe [110] edges of monolayer FeTe, as shown in Figs. 4(a) and 4(d). The obtained band structures and local density-of-states (LDOS) for bulk and edge atoms are displayed in Figs. 4(b)-4(c) for the [100] edge and Figs. 4(e)-4(f) for the [110] edge. In the calculations, the nontrivial gap is located about 0.2 eV above the Fermi level and gapless Dirac-cone-like edge states should appear in the region. Actually, this is the case for the (100) surface, as shown in Fig. 4(b). The edge states extend over nearly the whole Brillouin zone and merge into the bulk states near the M point. Due to the 1D nature of the Dirac band dispersion, the edge LDOS is almost a constant near the nontrivial gap. With decreasing energy, the edge LDOS increases rapidly and develops a peak at 0.2 eV below the gap, which is attributed to the flat dispersion of edge states. A similar peak can also be found in the LDOS of the [110] edge, although a constant LDOS nearly disappears due to the strong overlap between the edge and bulk states, as shown in Figs. 4(e)-4(f).

To demonstrate the edge states experimentally, *in situ* STM/STS measurements are carried out after growth of monolayer FeTe/STO. Topographic images of monolayer FeTe containing Fe-Te [100] and Fe-Fe [110] edges are presented in Figs. 4(g)-4(h), respectively. The insets of Figs. 4(g)-4(h) show the atomically-resolved topography indicating the high-quality of our sample. We take STS measurements along the lines indicated by the yellow arrows in Figs. 4(g) and 4(h) from the edges of the Fe-Te and Fe-Fe directions to the inside film, as shown in Figs. 4(i) and 4(j). The common feature is that noticeable enhancements around 50 meV below $E_F$ (30 meV below the η band bottom) appear for the edges compared with the bulk, which decay to the bulk value within several nanometers. According to the calculated results in Fig. 4(b), we find that the edge states connecting the η band at the Γ point and the electron δ band at the M point have a section of flat dispersion around the band bottom of the δ band, which probably corresponds to the enhanced DOS. Although the exact value of the calculated binding energy is not reliable due to strong correlations in FeTe, the band bottom of the δ band is reported experimentally to reside around 50 meV below $E_F$ [8], which is consistent with the energy observed in STM/STS measurements. Therefore, the increased DOS around this energy level is expected to be induced by topological edge states according to our LDA calculations. Furthermore, for the spectra of the Fe-Te edge, a plateau of DOS appears around 0 meV to 20 meV below $E_F$ in Fig. 4(i), which probably corresponds to the Dirac point of the edge states. The energy value of the Dirac point matches well with the band structure results shown in Fig. 2(c).

## IV. CONCLUSION

Our ARPES and LDA calculation results provide direct evidence of topological transition in high-$T_c$ FeTe$_{1-x}$Se$_x$/STO monolayers. The consistency of STM/STS experiments and ARPES/LDA results further supports it. The topological transition can be tuned by Te doping. Moreover, the position of topological bands can be tuned by abundant methods, including post anneal treatment [19], alkali metal doping [40, 41], changing various substrates [42-44] and electric field regulation [45-47]. To directly observe the topological edge states by ARPES, we suggest that FeTe$_{1-x}$Se$_x$ monolayers should be grown on STO substrates with densely parallel edges, which can be obtained by chamfering STO at specific angle.

FeTe$_{1-x}$Se$_x$/STO monolayers are the two-dimensional counterpart of FeTe$_{0.5}$Se$_{0.5}$ single crystal but with much higher $T_c$. Majorana bound states have been observed on the surface of the latter [11,12]. The superconductivity on topological edge states of FeTe$_{1-x}$Se$_x$/STO monolayers can be introduced by the bulk proximity effect. If a magnetic defect is deposited at the superconducting edges to suppress the superconductivity, Majorana bound states are expected to appear at the domain wall between the superconducting and magnetic insulating regions. Due to the high superconducting transition temperature in FeTe$_{1-x}$Se$_x$/STO monolayers [8, 28], our discovery of its topological nontrivial properties provides us a simple and tunable platform for realizing topological superconductivity and manipulating Majorana bound states at high temperature.

## ACKNOWLEDGMENTS

We thank P. Zhang, Z.-Q. Han and L.-Y. Kong for useful discussions. This work is supported by grants from the Ministry of Science and Technology of China (2016YFA0401000, 2015CB921000, 2016YFA0300600) and the National Natural Science Foundation of China (11574371, 11622435, U1832202, 11888101). Y.B.H. acknowledges support by the CAS Pioneer "Hundred Talents Program" (type C). P.R. acknowledges the funding from the Canada First Research Excellence Fund.

X.L.P., Y.L., H.B.D., H.Q.M. and Y.J.S. synthesized the samples. X.L.P. performed the ARPES measurements and analyzed the data with help from X.S., W.H.F., M.L. and Y.B.H.. Y.L., H.B.D. and H.Q.M. performed the STM/STS measurements and analyzed the data. X.X.W. and J.P.H. performed the LDA calculations and provided theoretical input. X.L.P., Y.L., X.X.W., P.R., J.P.H., H.Q.M., Y.J.S. and H.D. wrote the manuscript. All authors discussed the manuscript. H.D. and Y.J.S. supervised the project.

X.L.P., Y.L. and X.X.W. contributed equally to this work.

†Correspondence authors: mhq@iphy.ac.cn; yjsun@iphy.ac.cn; dingh@iphy.ac.cn

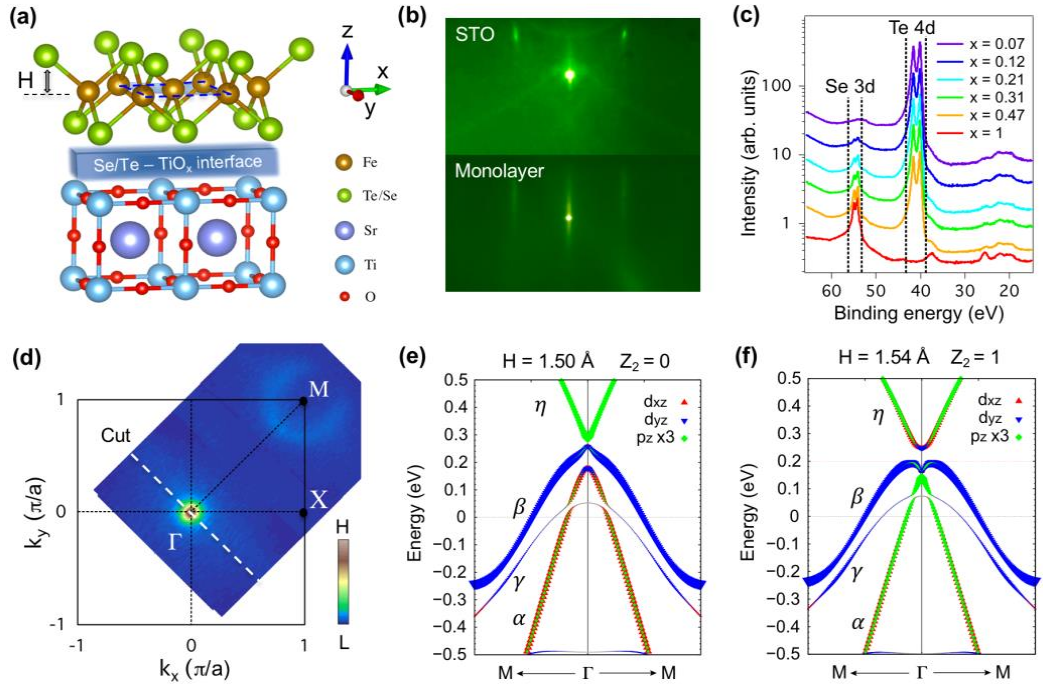

FIG. 1. (a) Schematic crystal structure of monolayer FeTe$_{1-x}$Se$_x$/STO. $H$ is defined as the Te/Se height relative to the Fe plane. (b) Typical RHEED pattern of STO substrate (upper panel) after treatment in vacuum and FeTe$_{1-x}$Se$_x$/STO monolayer (lower panel) after annealing. (c) *In situ* XPS results of six samples with different Se content *x* detected by 100 eV photons. (d) Definition of the Brillouin Zone of the two Fe unit cell. The Fermi surface of monolayer FeTe$_{0.81}$Se$_{0.19}$/STO is measured at T = 15 K. The intensity has been integrated in the ±10 meV energy range. White dashed line represents the cut position of ARPES spectrum shown as followed. (e)-(f) LDA calculation results of band structure for freestanding monolayer FeSe/STO with Se height $H$ = 1.50 Å and $H$ = 1.54 Å, respectively. Three hole-like bands (α, β, γ) and one electron-like band η are predicted around Γ. The main orbital of the γ band is d$_{xy}$ according to our calculation [Fig. S4 of Supplemental Material [38] ].

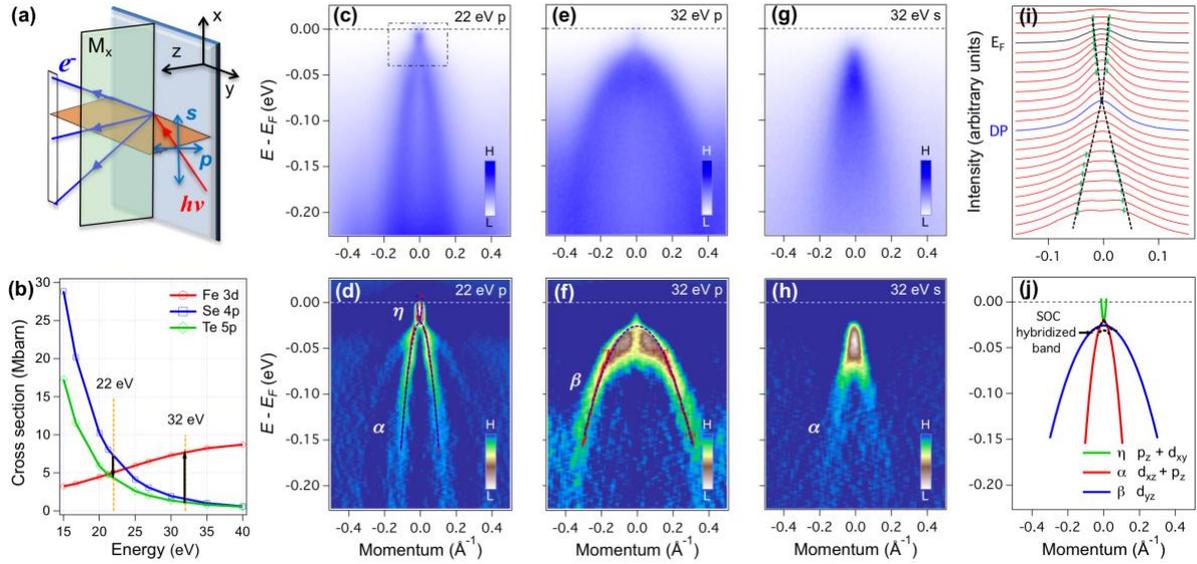

FIG. 2. Electronic structure of monolayer FeTe$_{0.79}$Se$_{0.21}$/STO around the Brillouin Zone center ($\Gamma$). (a) Schematic experimental geometry of ARPES measurements. (b) Comparison of cross section of Fe 3$d$ and Se 4$p$ (Te 5$p$) [37]. (c)-(d) Electronic structure around the $\Gamma$ point detected by 22 eV $p$-polarized photons and the corresponding second derivative spectrum. Red dots in (d) represent peak position of MDC fitted by Lorentz function. Blue dashed curves are fitting results of the MDC peak position, using the function $E = C_0 + C_1|k| + C_2 k^2$. (e)-(h) Same as (c)-(d) but measured with (e)-(f) 32 eV $p$-polarized photons and (g)-(h) 32 eV $s$-polarized photons, respectively. (i) MDC plot of the dashed rectangular region in (c). Green dots show MDC peak position fitted by Lorentz function. Black dashed lines are schematic of the Dirac-cone-like dispersion after tracing the position of MDC peaks. The MDCs at energies of the Fermi level ($E_F$) and the Dirac point (DP) are highlighted by black and blue, respectively. (j) Summary of the band structure and orbital analysis around $\Gamma$. The three solid curves are results of fitting of MDC peaks of the $\alpha$, $\beta$ and $\eta$ bands. Black dashed curves are the schematic results of band hybridization due to spin-orbit coupling (SOC).

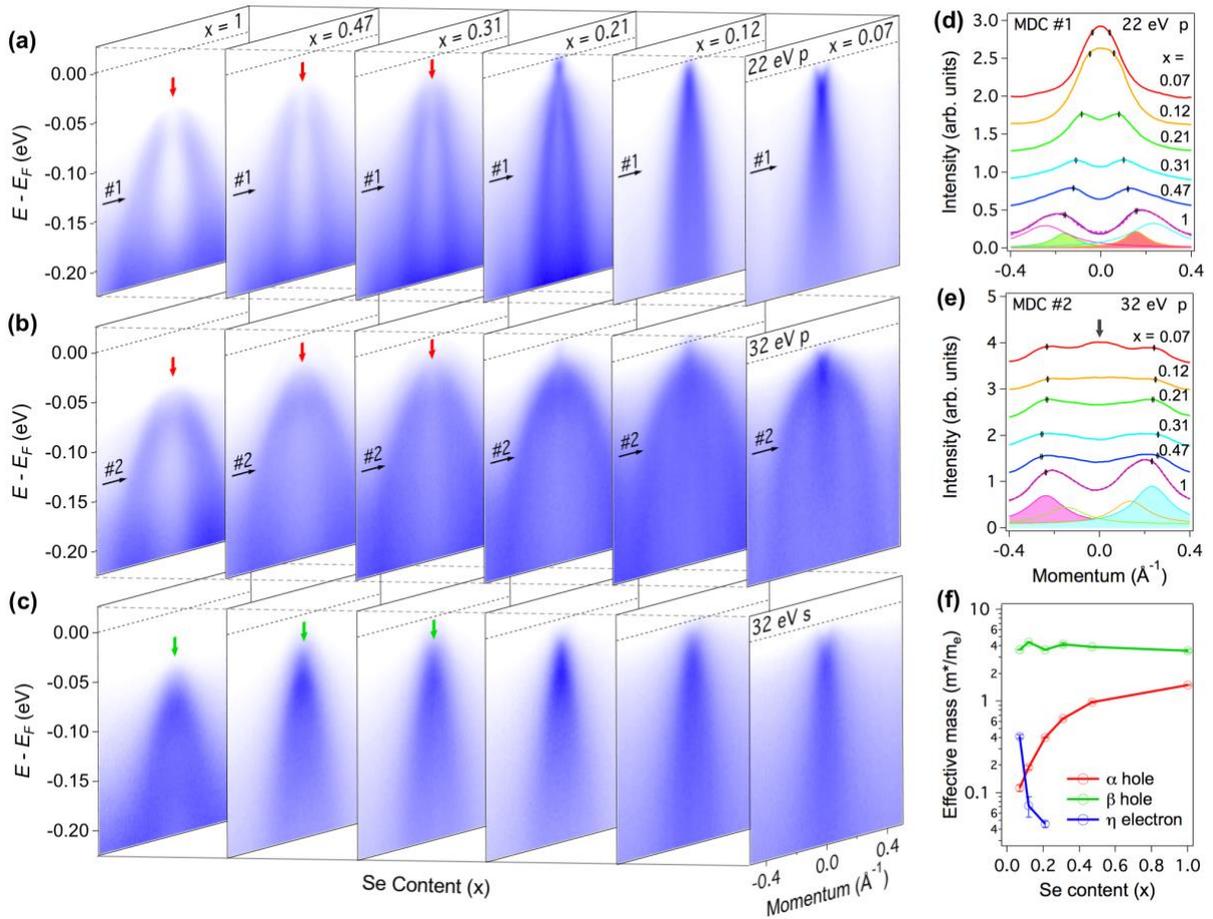

FIG. 3. Band inversion in monolayer FeTe$_{1-x}$Se$_x$/STO. (a)-(c) Electronic structure evolution with Se content, detected by (a) 22 eV *p*-polarized, (b) 32 eV *p*-polarized and (c) 32 eV *s*-polarized photons, respectively. The Se contents *x* of six samples (from left to right) are 1, 0.47, 0.31, 0.21, 0.12 and 0.07, which are determined by *in situ* XPS method. The red arrows in (a) and (b) show the relatively weak intensity around Γ before the topological transition. The green arrows in (c) indicate the appearance of band top. (d)-(e) MDC plot of six samples at energy positions pointed by #1 in (a) and #2 in (b), respectively. Position of #1 and #2 is 80 meV below the α and β band tops. Four-Lorentzian-function fitting results for the *x* = 1 sample are presented at the bottom of (d) and (e). Black dots on each MDC are fitting results of the peak positions. Lorentzian function curves associated with the α and β bands are shaded in (d) and (e), respectively. The black arrow in (e) indicates the appearance of the α band after the topological transition. (f) Effective mass evolution with Se content. The values of the α, β and η bands are shown by red, green and blue, respectively.

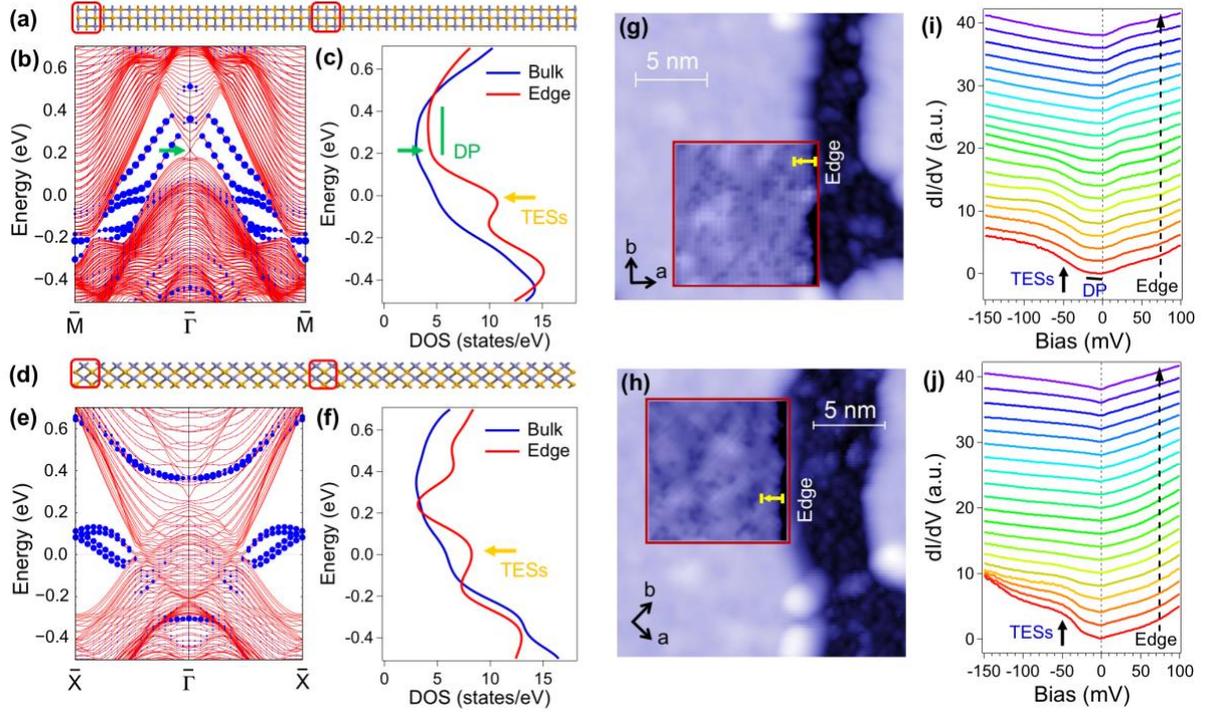

FIG. 4. Topological edge states in monolayer FeTe/STO. (a) Model of atomic chain along the Fe-Te direction. (b) One-dimensional band structure for the atomic chain in (a). The green arrow represents the position of the Dirac point (DP) and the sizes of the blue circles represent the contributions of the left edge. (c) Calculated local density-of-states (LDOS) of bulk and edge for model (a). (d)-(f) Same as (a)-(c) but for the atomic chain along the Fe-Fe direction. (g)-(h) Experimental STM Topography of the Fe-Te edge (V = 500 mV, I = 50 pA) and the Fe-Fe edge (V = 1 V, I = 50 pA) of FeTe/STO monolayer. Notice that the roughness of the film is mainly due to STO substrate instead of Te/Se height difference [Fig. S5 of Supplemental Material [38] ]. The inset shows the atomically resolved topography with the edge along Fe-Te (V = 100 mV, I = 0.5nA) and Fe-Fe direction (V = 100 mV, I = 1 nA), respectively. (i)-(j) dI/dV spectra (Junction setpoint V = 100mV, I = 0.2 nA) from edge to bulk detected along the yellow arrow shown in (g)-(h), respectively. The black arrows show the density-of-states enhancement near the edge. The plateau labelled DP in (i) indicates the existence of Dirac point.

| Pol. | $d_{xz}$ | $d_{yz}$ | $p_z$ | $d_{xy}$ |
|---|---|---|---|---|
| *s* ($A_x$) | ✓ | ✗ | k≠0 | ✗ |
| *p* ($A_y$) | ✗ | ✓ | ✗ | k≠0 |
| *p* ($A_z$) | k≠0 | ✗ | ✓ | ✗ |

TAB. 1.  ARPES matrix element effect of orbitals relevant to the band inversion process. "✓" indicates that the orbital can be observed by the corresponding polarized photons, while "✗" means that it cannot be observed. "k≠0" indicates that the orbital can be observed away from the Γ point.